# The Hubbard model and the absolute negative conductivity of graphene

**M B Belonenko[1], N G Lebedev[2], M M Shakirzyanov[3] and N N Yanyushkina[2]**

[1]Volgograd Institute of Business, Laboratory of Nanotechnologies, Volgograd, Russia

[2]Volgograd State University, Volgograd, Russia

[3]Kazan Physical-Technical Institute, Kazan, Russia

e-mail: yana_nn@inbox.ru

The analysis of Current-voltage characteristic and Ampere-Gauss characteristic for graphene with Hubbard interaction was carried out depending on frequency of external variable field and magnetic field. The regions of the absolute negative conductivity were found.

## 1. Introduction

Recently, graphene has attracted a great interest of many scientists. It is a structure which consists of one layer of carbon atoms, located in the units of hexagonal lattice. A great attention is paid to large electron mobility in the graphene and to its unique properties which are an alternative of silicic base in the modern microelectronics [1-4]. And a possibility of negative conductivity in non-equilibrium electron system, that is the current flows in the opposite direction to electric field, was specified by Krömer in the late of fifties [5]. But the first experimental works and experimental proofs of existence of an absolute negative conductivity (ANC) have appeared only in 2002 [6]. State of ANC is unstable. The system breaks up to domains, and measured macroscopical resistance becomes equal to zero. Thereby, effect of the absolute negative conductivity in graphene is a new object for researches, and this phenomenon is a very attractive for the experimental appendices.

We note that electromagnetic waves in the carbon structures become strongly nonlinear even in the weak fields that gives rise to spread possibility of solitary electromagnetic waves in the graphene and carbon nanotubes. These properties of carbon nanostructures have theoretical interest and attempts of applying in the nonlinear optics [7]. The questions are associated with the nonlinear response carbon nanotubes on the electromagnetic field were considered in [8, 9]. Nonlinearity is caused by change of classical function of electron distribution and by non-parabolic dispersion law of electrons. Possibility of soliton existence and dependence of their parameters on carbon nanotube parameters has been established in [10-12]. It is also very important, that one of the main factors in these works was a description of electron subsystem in the frameworks of Boltzmann kinetic equation in approach of constant relaxation time. That demands a microscopic substantiation. The electronic properties themselves are often left outside of the consideration. Nevertheless, namely these properties can be exhibited in the optical part of the spectrum. For example, Coulomb interaction between electrons can lead to the changes in the dispersion law and the corresponding changes in the optical response of the structures. It is worth noting that it seems the simplest way to consider the Coulomb interaction is to adopt the Hubbard model [13, 14], where the Coulomb repulsion only of the electrons localized at the same lattice site is taken into account. The situation when we can operate the important characteristics of electron subsystem (including conductivity) by operating a spectrum of charge carries. From the physical point of view, the reason consists in the phenomenon similar to the Hall effect is a deviation moving of electromagnetic field electrons under the influence of the electromagnetic field pulse. It should be noted, that recently, the interest to this problem has risen [15].

Summarizing, one can draw a conclusion, that the problem of graphene response in the magnetic field with taking into account Hubbard interaction, is very important and actual.

## 2. Basic equations

Let us consider the response of graphene on external electric field along axis x in geometry is when magnetic field is perpendicular to graphene mnolayers.

Then the Hamiltonian of the electron system can be written in the Hubbard form [13, 14]:

$$H = H_0 + H_{\text{int}}$$

$$H_0 = \sum_{j\Delta\sigma} t_0 a_{j\sigma}^+ a_{j+\Delta\sigma} + h.c.$$

$$H_{\text{int}} = U \sum_j a_{j\sigma}^+ a_{j\sigma} a_{j-\sigma}^+ a_{j-\sigma}$$

where $a_{j\sigma}^+$, $a_{j\sigma}$ are the creation and annihilation operators of electrons in a graphene unit with number j and spin $\sigma$, $t_0$ is the overlap integral between adjacent grapheme units determined by overlapping of the wave functions of the grapheme electrons, $\Delta$ is the vector relating two adjacent units in the hexagonal, $U$ is the energy of Coulomb repulsion of the electrons at the same unit.

Using the Fourier transform:

$$a_{n\sigma}^+ = \frac{1}{N^{1/2}} \sum_j a_{j\sigma}^+ \exp(ijn)$$

$$a_{n\sigma} = \frac{1}{N^{1/2}} \sum_j a_{j\sigma} \exp(-ijn)$$

(1)

which diagonalizes the Hamiltonian $H_0$, it is easy to obtain the electron spectrum of the subsystem, in the case where the Coulomb repulsion is absent $\varepsilon(p)$.

The interaction Hamiltonian $H_{int}$ was considered in different approximations and various approach; there is a lot of publications on this topic, we mention here only [15-17]. The main result following from the analysis carried out in these works consists in the statement that, if the term with $U$ is taken into account, the spectrum of elementary excitations used in the model changes. So, two bands originally degenerated in the spin projections $\sigma$ are split in two nondegenerated bands with the spectrum described in [15-17] as follows:

$$E(\vec{p}) = \varepsilon(\vec{p})/2 + U/2 \mp \sqrt{\varepsilon^2(\vec{p}) - 2\varepsilon(\vec{p})U(1 - 2n_0) + U^2}/2 \qquad (2)$$

where $\varepsilon(\vec{p})$ is given by (3), $n_0$ is the number of electrons in the unit.

The physical reason of a change of the spectrum of elementary excitations is obvious enough, being explained by the electron scattering on the fluctuations of the Coulomb field created by the electrons with different spin (described by $H_{int}$). It should be also noted, that a spectrum similar to (2) can be written under the conditions of electron-electron and electron-phonon interactions, if $t_0$ is understood as corresponding renormalized constant.

The seed spectrum of electrons in graphene without taking into account the Coulomb interaction of electrons at the same unit is known to be of the [18]:

$$\varepsilon(\vec{p}) = \pm\gamma\sqrt{1 + 4\cos(ap_x)\cos(ap_y/\sqrt{3}) + 4\cos^2(ap_y/\sqrt{3})} \qquad (3)$$

where $\gamma \approx 2.7$ eV, $a = 3b/2\hbar$, $b = 0.142$ nm is the distance between adjacent carbon atoms in graphene, $\vec{p} = (p_x, p_y)$. Different signs are related to conductivity band and to valence band.

In order to determine the current, let us use a semiclassical approximation [19] that takes the dispersion law (2) from the quantum-mechanical model and describes evolution of the ensemble of particles by the classical Boltzmann kinetic equation in the relaxation time approximation as follows:

$$\frac{\partial f}{\partial t} + (-\frac{q}{c}\frac{\partial A_x}{\partial t} + qhv_y)\frac{\partial f}{\partial p_x} + (-\frac{q}{c}\frac{\partial A_y}{\partial t} - qhv_x)\frac{\partial f}{\partial p_y} = \frac{F_0 - f}{\tau} \qquad (4)$$

Here, $h$ is the external magnetic field parallel to the z axis, and choosing: $\vec{E} = -\frac{1}{c}\frac{\partial \vec{A}}{\partial t}$, where $\vec{A} = (A_x, 0, 0)$, $v_x = \partial E / \partial p_x$, $v_y = \partial E / \partial p_y$, $F_0$ is the equilibrium Fermi distribution function written as:

$$F_0 = \frac{1}{1 + \exp(E(\vec{p})/k_b T)},$$

$T$ is the absolute temperature, $k_b$ is the Boltzmann constant. $\tau$ is the relaxation time that can be evaluated according to [18] as approximately $10^{-12}$ s. Since the typical ultrashort light pulse duration is on the order of $10^{-15}$ s. Let us write the current components $j_\alpha(z,t)$ ($\alpha = x, y$):

$$j_\alpha = q \int dp v_\alpha f \tag{5}$$

Use the average electron method [20-21], according to which the current can be expressed via a solution of the classical equation of motion for an electron in present fields $A_x = A_0 \cos(\omega t)$ as:

$$\frac{dp_x}{dt} = qA_0 \cos(\omega t) - qhv_y$$
$$\frac{dp_y}{dt} = qhv_x \tag{6}$$

With the initial conditions: $p_x|_{t=0} = p_{x0}$; $p_y|_{t=0} = p_{y0}$. In the limit of low temperatures, this method leads to the following relation:

$$j_\alpha = q \int e^{-t} v_\alpha(p_x, p_y) dt \tag{7}$$

where $p_x, p_y$ are the solutions of (5) with some initial conditions. Here, the condition of low temperatures implies the validity of inequality: $k_B T << \Delta$, which eliminates averaging with respect to initial momentum in the average electron method. It is worth noting, that a high temperatures, formula (7) should be used with a solution of (6) for arbitrary initial conditions, followed by averaging with an equilibrium distribution function in which the role of momentum is played by the initial conditions for $p_x, p_y$.

Note, $p_{x0} = p_{y0} = 0$ are initial conditions in a classical variant of average electron method, that was associated with the least energy $\varepsilon(p)$ in [20-22] accorded to a center of Brilluene. In our case, it is necessary to consider the solutions of (6) for four initial conditions (correspond to minimum of $E(p)$): $ap_x = 0; ap_x = \pi; ap_y = \pi/3; ap_y = 2\pi/3$. Then we should sum up all values for current. Also we should note that since we consider low temperature case, then in average electron method we have to take into account only initial conditions which conform to energy minimum. It does not take into account energy, which is associated with spin levels of electrons in a magnetic field $\pm \hbar h/2$, since it is chosen the distance with the least energy and this addition has not a big influence on velocities $v_x, v_y$.

In the general case, where the dispersion law is set by (2), it is difficult to obtain a solution valid in the entire range of alternating electric and magnetic fields. For the cosines and parabolic dispersion laws, the possible solutions were reported in [20, 22]. Since in this case it is difficult to receive such solution, (6) were solved numerically without any limits on the values of electric and magnetic fields.

**3. Results of Numerical Calculations**

The equations under study were solved numerically, using a Runge-Kutt method with the eighth order of accuracy with following initial conditions: $ap_x = 0; ap_x = \pi; ap_y = \pi / 3; ap_y = 2\pi / 3$.

The current-voltage characteristic for the frequency $\omega = 1 / \tau$ and for different values of magnetic field is given in Fig.1.

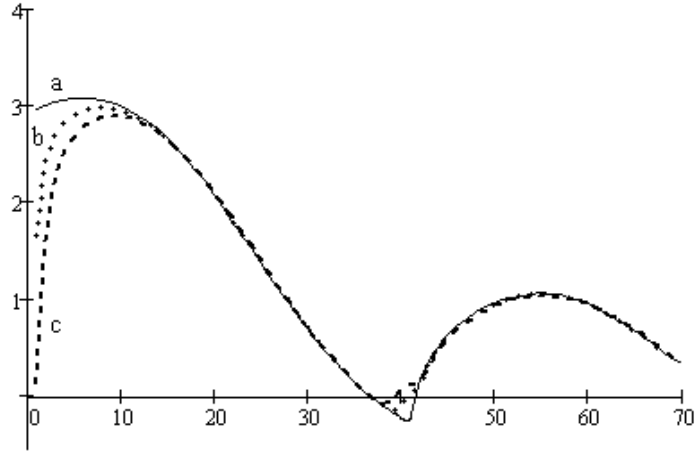

Fig. 1. Current-voltage characteristic for the graphene in a case of different magnetic fields. For b) magnetic field is in 3 times lager then for a), and for c) magnetic field is in 5 times lager then for a). The x axis is a dimensionless strength of the electric field (one unit corresponds to $10^6$ V/m), the y axis is a current (in dimensionless units).

It can be seen that there is a region with an absolute negative conductivity in the current-voltage characteristic, besides ordinary region with a differential negative conductivity, which is peculiar all substances with periodical dispersion law. It should be associated with non-equilibrium state of the system of graphene electron, at first caused by the strong non-parabolic dispersion law. Also we note that such state will be unstable and will lead to graphene segmentation into domains. The region of absolute negative conductivity decreases when magnetic field increases. It is associated with a general decreasing of the current at the increasing of the magnetic field. It can be quality understood when we consider classical action of the Lorentz power, which declines the electrons propagating in the x axis.

Ampere-Gauss characteristic for the frequency $\omega = 1 / \tau$ and for different values of variable electric field is illustrated in Fig.2.

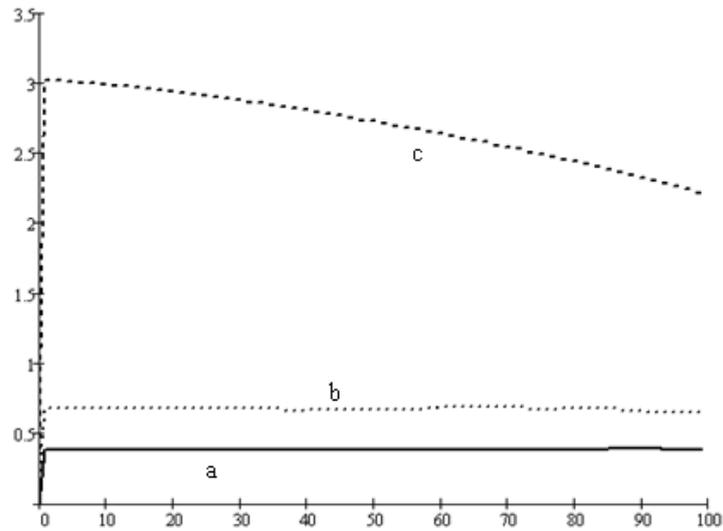

Fig. 2. Ampere-Gauss characteristic for the graphene in a case of different variable electric fields. For b) amplitude of electric field is in 3 times lager then for a), and for c) amplitude of electric field is in 10 times smaller then for a). The x axis is a value of constant magnetic field (in dimensionless units), the y axis is a current (in dimensionless units).

It can be seen from given dependence that magnetizing force has a weak influence on the current, and the value of current is determined by the variable electric field. That is coordinates with Fig. 1. It is worth noting, that the dependence current on amplitude of the variable electric field is not monotone. That is also coordinates with the giving data.

Dependence of the current on the frequency of the external variable electric field when the magnetic field is a constant, and for different amplitudes of the variable external electric field is given in Fig. 3.

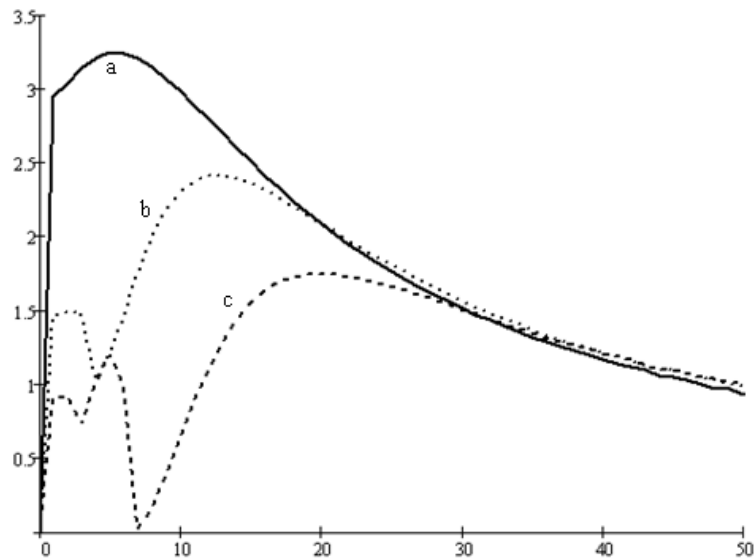

Рис. 3. Dependence of current on a frequency in a case of different variable electric fields. For b) amplitude of electric field is in 3 times lager then for a), and for c) amplitude of electric field is in 5 times lager then for a). The x axis is a frequency (in 1/τ units), the y axis is a current (in dimensionless units).

According to given data, there is a state with zero conductivity in grapheme, which appears with increasing of the amplitude of the variable electric field. We suppose, this phenomena has the same mechanism as appearance of regions with the absolute negative conductivity in the current-voltage characteristic. There are big possibilities for practical applications, because of this state is stable. Thereby, the graphene conductivity in the magnetic field can be analysed, using the average electron method without any limits on amplitude of the applying fields. The revealing states with the absolute negative conductivity will lead to division of the graphene sheet into domains, and will lead to appearance of the stable states with zero resistance.

## 4. Conclusion

In conclusion we formulate our main results:

1. A method for calculation of current-voltage characteristic and Ampere-Gauss characteristic in a case of strong electric and magnetic fields was offered.

2. There are regions of an absolute negative conductivity in graphene in a case of constant magnetic field, which is perpendicular to the graphene plane.

3. A conductivity of the graphene strongly depends on frequency of the external electric field in a case of magnetic field. And, in particular, the states with zero conductivity are possible.

4. The graphene conductivity is defined by amplitude of the external variable field and it has a weak dependence on amplitude of the applying constant field.

## Acknowledgments

This work was supported by the Russian Foundation for Basic Research under project No. 08-02-00663 and by the Federal Target Program "Scientific and pedagogical manpower" for 2010-2013 (project № NK-16(3)).